\documentclass[twocolumn]{aastex631}
\usepackage{xcolor}
\usepackage{amssymb, amsmath, apjfonts, natbib, color} 
\usepackage{rotating, graphics}
\usepackage{CJK}
\usepackage{apjfonts}
\usepackage{amsmath}

\usepackage{color}

\newcommand{\nup}{\nu_{\mathrm{p}}}
\newcommand{\lognup}{\log\nu_{\mathrm{p}}}

\newcommand{\lognupfnup}{\log\nu_{\mathrm{p}}f_{\nu_{\mathrm{p}}}}
\newcommand{\logf}{\text{log\_{f}}}

\def\degree{{\circ}} 
\newdimen\digitwidth 
\setbox1=\hbox{0} 
\digitwidth=\wd1 
\catcode`"=\active 
\def"{\kern\digitwidth} 

\begin{document} 
\title{Spectral Energy Distribution Variability of the Blazar OJ 287 during 2009-2021}

\author[0000-0002-4521-6281]{Wenwen Zuo}
\affiliation{Shanghai Astronomical Observatory, Chinese Academy of Sciences, 80 Nandan Road, Shanghai 200030, People's Republic of China}
\email{wenwenzuo@shao.ac.cn (WWZ), acgupta30@gmail.com (ACG)}

\author[0000-0002-9331-4388]{Alok C. Gupta}
\affiliation{Aryabhatta Research Institute of Observational Sciences (ARIES), Manora Peak, Nainital 263001, India}
\affiliation{Shanghai Astronomical Observatory, Chinese Academy of Sciences, 80 Nandan Road, Shanghai 200030, People's Republic of China}

\author[0000-0002-4455-6946]{Minfeng Gu}
\affiliation{Shanghai Astronomical Observatory, Chinese Academy of Sciences, 80 Nandan Road, Shanghai 200030, People's Republic of China}

\author[0000-0001-8580-8874]{Mauri J. Valtonen}\thanks{visitor}
\affiliation{Institute of Astronomy, University of Cambridge, Madingley Road, Cambridge CB3 0HA, UK}

\author[0000-0001-6158-1708]{Svetlana G. Jorstad}
\affiliation{Institute for Astrophysical Research, Boston University, 725 Commonwealth Avenue, Boston, MA 02215, USA}
\affiliation{Saint Petersburg State University, 7/9 Universitetskaya nab., 199034 St. Petersburg, Russia}

\author[0000-0003-2483-2103]{Margo F. Aller}
\affiliation{Astronomy Department, University of Michigan, Ann Arbor, MI 48109, USA}

\author[0000-0002-0393-0647]{Anne L\"ahteenm\"aki} 
\affiliation{Aalto University Mets\"ahovi Radio Observatory, Mets\"ahovintie 114, 02540 Kylm\"al\"a, Finland}
\affiliation{Aalto University Department of Electronics and Nanoengineering, P.O. BOX 15500, FI-00076 Aalto, Finland}

\author[0000-0001-6314-9177]{Sebastian Kiehlmann}
\affiliation{Institute of Astrophysics, Foundation for Research and Technology-Hellas, GR-71110 Heraklion, Greece}

\author[0000-0001-6890-2236]{Pankaj Kushwaha}
\affiliation{Department of Physical Sciences, Indian Institute of Science Education and Research Mohali, Knowledge City, Sector 81, SAS Nagar, Punjab 140306, India}

\author[0000-0003-1945-1840]{Hugh D. Aller}
\affiliation{Astronomy Department, University of Michigan, Ann Arbor, MI 48109, USA}

\author[0000-0002-1908-0536]{Liang Chen}
\affiliation{Shanghai Astronomical Observatory, Chinese Academy of Sciences, 80 Nandan Road, Shanghai 200030, People's Republic of China}

\author[0000-0001-9152-961X]{Anthony C. S. Readhead}
\affiliation{Owens Valley Radio Observatory, California Institute of Technology, Pasadena, CA 91125, USA}

\author[0000-0003-1249-6026]{Merja Tornikoski} 
\affiliation{Aalto University Mets\"ahovi Radio Observatory, Mets\"ahovintie 114, 02540 Kylm\"al\"a, Finland}

\author[0000-0003-4671-1740]{Qi Yuan}
\affiliation{Changchun Observatory, National Astronomical Observatories, Chinese Academy of Sciences, West Hill of Jingyue Lake, Changchun 130117, People's Republic of China}

\begin{abstract}
Using nearly simultaneous radio, near-infrared, optical, and ultraviolet data collected since 2009, we constructed 106 spectral energy distributions (SEDs) of the blazar OJ 287. These SEDs were well-fitted by a log-parabolic model.
By classifying the data into `flare' and `quiescent' segments, we found that the median flux at peak frequency of the SEDs during flare segments was 0.37$\pm$0.22 dex higher  compared to quiescent segments, while no significant differences were observed in the median values of the curvature parameter $b$ or the peak frequency $\lognup$. A significant bluer-when-brighter trend was confirmed through a relation between $V$ magnitude and $B-V$ color index, with this trend being stronger in the flare segments. Additionally, a significant anti-correlation was detected between $\lognup$ and $b$, with a slope of 5.79 in the relation between $1/b$ and $\lognup$, closer to the prediction from a statistical acceleration model other than a stochastic acceleration interpretation, though a notable discrepancy persists. This discrepancy indicates that additional factors, such as deviations from idealized conditions or radiative contributions—such as thermal emission from the accretion disk in the optical-UV range during quiescent states—may play a role in producing the observed steeper slope. Within the framework of statistical acceleration mechanism, lack of correlation between change in peak intensity and change in peak frequency suggests that change in electron energy distribution is unlikely to be responsible for the time-dependent SED changes. Instead, changes in Doppler boosting or magnetic fields may have a greater influence.
\end{abstract}


\section{INTRODUCTION}
Blazars are a subclass of radio-loud active galactic nuclei. They are further divided into two subclasses: flat-spectrum radio quasars (FSRQs) with strong emission lines \citep[e.g.][]{1978PhyS...17..265B,1997A&A...327...61G} and BL Lacertae objects (BL Lacs), which have either no emission lines or very weak (Equivalent width EW < 5 \AA) emission lines \citep{1991ApJS...76..813S,1996MNRAS.281..425M}. High brightness, high polarization, and extremely variable emission which is mostly non-thermal, spanning the whole electromagnetic (EM) spectrum are the main characteristics of blazars. Typically, the emission is ascribed to the relativistic jet that is pointed near the line of sight (LOS) of the observer \citep{1995PASP..107..803U}. Their multi-wavelength (MW) spectral energy distribution (SED) is a double-humped structure. The low-energy hump, which is caused by synchrotron emission from non-thermal electrons in the jet, peaks somewhere in the infrared (IR) to soft X-ray energy range, while the high-energy hump peaks in GeV to TeV $\gamma-$ray energies and is likely caused by inverse Compton (IC) up-scattering of synchrotron (SSC, synchrotron self-Compton) or external photons (EC, external Compton) by the relativistic electrons responsible for producing the synchrotron emission \citep{1998A&A...333..452K,2010ApJ...718..279G}.  

Blazars are one of the best examples of persistent, highly variable, but non-catastrophic sources in the era of MW transient astronomy. Studying changes in flux variability of blazars is a valuable way to uncover the physical processes behind the source's various states—whether low, high, or during outbursts. Simultaneous MW studies have been carried out in order to understand their emission mechanism spanning the whole EM spectrum \citep[e.g.][and references therein]{1997ApJ...486..799U,2005A&A...442..895A,2009ApJ...696L.150A,2007A&A...473..819R,2008A&A...485L..17R,2015MNRAS.454..353R,2009ApJ...690.1018V,2010ApJ...712..405V,2009A&A...504L...9V,2017MNRAS.472..788G,2018MNRAS.473.1145K,2018ApJ...863..175G,2020MNRAS.498L..35K,2024MNRAS.529.3894M}.  

The BL Lac OJ 287 ($\alpha_{\rm 2000.0}=08^{h}54^{m}48.^{s}87$, $\delta_{\rm 2000.0}=+20^{\degree}06'30.''64$) is at redshift $z=0.306$ \citep{1985PASP...97.1158S}. OJ 287 has been observed in optical bands since 1888 \citep{2024ApJ...968L..17V}. A small fraction of the light curve was available already in 1982 when it was noticed that OJ 287 may exhibit a nearly periodic outburst about every 12 years. The next outburst was expected in 1983 and it was indeed detected \citep{1988ApJ...325..628S}. The authors postulated a supermassive binary black hole (SMBBH) model to explain the 12 yr periodicity and predicted that the next outburst would take place in late 1994. \citet{1988ApJ...325..628S} also noted a possible shorter periodicity in the fades, the times of minimum light. Assuming that the difference in the periodicities arises from the procession of the major axis of the binary, \citet{1988ApJ...325..628S} calculated that the primary black hole's mass was $\sim$ 5 $\times \rm{10}^{9} \rm{M}_{\odot}$, while the secondary's mass was estimated from the rapid variability in 15.7 min timescale as $\sim$ 2 $\times \rm{10}^{7}\ \rm{M}_{\odot}$ \citep{1985Natur.314..148V}. The anticipated outburst was observed in 1994 thanks to a global optical monitoring campaign of the source known as OJ-94 \citep{1996A&A...305L..17S}. However, \citet{1996ApJ...460..207L} predicted that the outbursts should have a double peaked structure and that second peak should take place within a two-week interval in October 1995. It was immediately verified by observations \citep{1996A&A...315L..13S}. 

\citet{1997ApJ...484..180S} calculated the binary model forward to predict the next pair of outbursts in November 2005 and September 2007. The increase of the two-flare interval is due to orbit procession in the model and it improved the primary mass to $\sim$ 1.7 $\times \rm{10}^{10}\ \rm{M}_{\odot}$. Both flares were seen at expected times \citep{2011AcPol..51f..76V}. Another set of flares, this times a triple set, was predicted for years 2015, 2019 and 2022 \citep{1997ApJ...484..180S}. The model showed that the timing of the first flare was sensitive to the spin value of the primary. After it was observed, the spin value was calculated \citep{2016ApJ...819L..37V}. The timing of the second flare was very precise \citep{2020ApJ...894L...1L}. \citet{2018ApJ...866...11D} have developed a highly accurate SMBBH model that can forecast the time of the flares to within four hours. 
The last of the triple flares was not observable from the ground since it was expected when OJ 287 is very close to the sun. However, it was possible to infer the presence of the third flare from particular pre-flare activity \citep{2023MNRAS.521.6143V}. The BH binary model of \citet{2018ApJ...866...11D} yields the following values for OJ 287: primary BH mass = (18.35 $\pm$ 0.05) $\times \rm{10}^{9}\ \rm{M}_{\odot}$, and secondary BH mass =  (150 $\pm$ 10) $\times \rm{10}^{6}\ \rm{M}_{\odot}$.

There are many claims of detections of QPOs from OJ 287 on a wide variety of timescales, from a few tens of minutes to decades and more over multiple EM bands, aside from the well-established 12-year and 55-year periodicities in the optical band \citep{2006ApJ...646...36V}. \citet{1973ApJ...179..721V} reported for the first time the detection of $\sim$40-min optical QPO in OJ 287 using accurate optical photoelectric observations on March 18, 1972. Later, a few more optical QPOs were reported with periods ranging from 23 to 40 minutes \citep{1974MNRAS.168..417F,1985Natur.314..146C}. In April 1981 observation of the source in 37 GHz radio band, a $\sim$15.7 minutes QPO was reported \citep{1985Natur.314..148V}. Using recent advance techniques, there are more claims of detection of QPOs in OJ 287 in different EM bands on diverse timescales ranging from a few tens of days to months to years in the different time spans of data \citep[e.g.][and references therein]{2013MNRAS.434.3122P,2016ApJ...832...47B,2018MNRAS.478.3199B,2020MNRAS.499..653K}.   

OJ 287 has been observed simultaneously on various flux, spectral, and polarization states on several occasions with diverse timescales \citep[e.g.][and references therein]{2017MNRAS.468..426S,2018MNRAS.473.1145K,2018MNRAS.479.1672K,2020Galax...8...15K,2018ApJ...863..175G,2020MNRAS.498L..35K,2021ApJ...923...51K,2021A&A...654A..38P}. The source has shown a major $\gamma-$ray flare in Fermi observation in 2009, which was studied to understand the high energy emission mechanism during this episode \citep{2013MNRAS.433.2380K}. Extensive X-ray flux and spectral variability of OJ 287 have been studied on several occasions using various X-ray and MW space missions, and variabilities have been found on diverse timescales \citep[e.g.][and references therein]{1997PASJ...49..631I,2001PASJ...53...79I,2018MNRAS.479.1672K,2018MNRAS.480..407K,2020ApJ...890...47P,2021MNRAS.504.5575K,2021ApJ...923...51K,2022MNRAS.509.2696S,2022MNRAS.510.5280M,2024MNRAS.532.3285Z}. 

When observing blazars at multiple epochs, simultaneous multi-wavelength SEDs provide valuable information about the emission mechanism of blazars in their various flux levels \citep[e.g.][and references therein]{1996ApJ...463..444S,2006A&A...445..441N,2004A&A...413..489M,2006A&A...448..861M,2008A&A...478..395M,2011MNRAS.417.1881R,2014MNRAS.444.3647B,2021MNRAS.504.5074S,2022MNRAS.513.4645S,2022MNRAS.517.2757S}. Modeling the broad-band SEDs of blazars is essential to understand the extreme conditions within different emission regions. This approach helps us comprehend the dynamic phenomena shaping the observed behavior of blazars. In the ideal case, such studies require simultaneous data in multiple bands. In the present paper, by utilizing comprehensive data spanning radio, near infrared (NIR), optical, and ultraviolet (UV) bands for OJ 287, we construct multi-epoch flux state-specific SEDs from nearly simultaneous observations, strictly maintaining temporal intervals of up to 10 days.

We describe the observations and data in Section \ref{sec:data}, SED modeling in Section \ref{sec:sed_modelling}. The results are delivered in Section \ref{sec:results} and discussed in Section \ref{sec:discussion}. We summarize our main results in Section \ref{sec:conclusion}. Throughout the paper, a flat $\Lambda$CDM cosmology with $\Omega_{\Lambda} = 0.7$, $\Omega_{m} = 0.3$ and $H_{0} = \rm 70\ km\ s^{-1}\ Mpc^{-1}$ is adopted.

\section{Observations and Data\label{sec:data}}
Multi-band radio, NIR, optical, and UV data of the blazar OJ 287 are collected for the period of 1998 - 2023 from various public archives and observing facilities. The details about the data are provided in Table~\ref{table:obslog}.  

The UVOT is one of the instruments onboard the Swift observatory, capable of observing in six filters namely $V$, $B$, $U$, $w1$, $m2$, $w2$ covering optical to UV regions of the EM spectrum. We used all the observation IDs from 2005 to 2023 and analyzed following the standard data reduction prescription as mentioned in \cite{2021ApJ...921...18K} and \cite{2023arXiv230516144K}. 

Optical $V$ and $R$ band photometric observations of OJ 287 are obtained from the spectro-polarimeter mounted on the 2.3 m Bok and 1.54 m Kuiper telescopes at Steward Observatory, University of Arizona, USA. OJ 287 data from October 2008 to June 2018 are taken from the public archive\footnote{http://james.as.arizona.edu/$\sim$psmith/Fermi/DATA/Objects/} of the Steward Observatory. The details of the instrument, observational program, observations and the data analysis procedures are provided in details in \citet{2009arXiv0912.3621S}. 

Optical $B$, $V$, $R$, and $I$ bands photometric observations of OJ 287 were carried out from January 2006 to February 2023 at the Perkins telescope of the Perkins Telescope Observatory (Flagstaff, AZ, USA). The details about the instrument, observations and the data analysis methods are given in \citet{2010ApJ...715..362J}. 

Optical $B$, $V$ and $R$ band data, and NIR J, K band data for OJ 287 are taken from the public archive of the SMARTS (Small and Moderate Aperture Research Telescope System)\footnote{http://www.astro.yale.edu/smarts/glast/home.php} from February 2008 to April 2017. SMARTS consist of 0.9 m, 1.0 m, 1.3 m, and 1.5 m telescopes at the Cerro Tololo Inter-American Observatory (CTIO) in Chile. These telescopes observed the blazars at both NIR and optical wavelengths that Fermi-LAT monitors. The SMARTS telescopes, detectors, observations, and data analysis details are provided in \cite{2012ApJ...756...13B} and \cite{2012AJ....143..130B}. 

The $J$, $H$ and $K_{s}$ NIR band observations of OJ 287 from October 1995 to November 2021
were carried out with 2.12 m telescope which is equipped with a NIR camera named CANICA (Cananea Near-Infrared Camera) of the Guillermo Haro Astrophysical Observatory (OAGH) located in Cananea, Sonora, Mexico. The details of the instrument, observations and the data analysis procedures are provided in details \citep[e.g.][]{1989ApJ...345..245C,2017RMxAA..53..497C,2022ApJS..260...39G}, and also the photometric data is already published in \cite{2022ApJS..260...39G}. 

The UMRAO (University of Michigan Radio Astronomy Observatory) flux density data of OJ 287 at 4.8, 8.0 and 14.5 GHz from November 2007 to June 2012 are obtained from the Michigan 26-m equatorially mounted, prime focus, paraboloid as part of the University of Michigan extra-galactic variable source monitoring program \citep{1985ApJS...59..513A}. The radio data of OJ 287 at 15 GHz is taken from the blazar monitoring program of 40m telescope of Owens Valley Radio Observatory (OVRO) for the period of 2008 January to 2023 August. The details of this observational program, observations and the data analysis procedures are provided in \citet{2011ApJS..194...29R}. 

Using the 14-meter radio telescope at Aalto University Mets\"ahovi Radio Observatory in Finland, observations of OJ 287 at 37.0 GHz were conducted. \citet{1998A&AS..132..305T} provided a thorough explanation of the Mets\"ahovi data reduction and analysis process. 

The VLBA-BU BLAZAR monitoring effort involves about monthly VLBA observations of a sample of AGNs identified as gamma-ray sources at 43 GHz and 86 GHz. The observations and data analysis of OJ 287 at 43 GHz and 86 GHz 
are presented in detail \citep[][and references therein]{2017ApJ...846...98J,2022ApJS..260...12W}. 
OJ 287 is a very compact core-dominated source at radio wavelengths, especially at high radio frequencies such as 43 and 86 GHz. As described in \cite{2017ApJ...846...98J}, for each epoch we calculated the total flux density in the images of several sources in the sample that are known to have very weak emission outside the angular size range of the VLBA images (0235$+$164, 0420$-$014, 0716$+$714, OJ 287, and 1156$+$295). These values were compared with total flux densities obtained by interpolating in time the measurements of these sources by monitoring programs carried out at the VLA\footnote{http://www.vla.nrao.edu/astro/calib/polar/} and the Effelsberg telescope at 43 GHz and the POLAMI program at 86 GHz. The comparison produced the flux-density correction factors, which in general are of order 1.1-1.3 at 43 GHz but can reach values of 2-3 at 86 GHz. The factors were applied for final adjustment of the flux-density scale in the images. Therefore, these correction factors take care of the extended structure of OJ 287 outside the VLBA scale and give estimates of uncertainties of flux density values at 43 GHz  $\sim$10\% and at 86 GHz $\sim$15\%.

\begin{deluxetable*}{cccccc}
    \tabletypesize{\footnotesize}
    \tablecaption{Radio, NIR, Optical and UV bands observation log of OJ 287 \label{table:obslog}}
    \tablewidth{0pt}
    \tablehead{\colhead{Observatory} & \colhead{Bands} & \colhead{Duration}  & \colhead{$MJD$ Duration Applied} &  \colhead{Duration Applied}  &  \colhead{$N_{\rm data}$}  \\
          \colhead{(1)} & \colhead{(2)} & \colhead{(3)} & \colhead{(4)}  & \colhead{(5)}  & \colhead{(6)} \\}  
    \startdata
  SWIFT                    & $w2$, $m2$, $w1$    &  2005-05-20 to 2023-01-20  &  54850 to 59225  &  2009-01-19 to 2021-01-11 &  616, 589, 626 \\
                           & $U$, $B$          &  2005-05-27 to 2023-01-20  &  54850 to 59224  &  2009-01-19 to 2021-01-10 &  594, 580  \\
                           & $V$             &  2005-05-20 to 2023-01-20  &  54850 to 59225  &  2009-01-19 to 2021-01-11 &  561  \\
  Steward Observatory      & $V$, $R$          &  2008-10-04 to 2018-06-23  &  54850 to 58292  &  2009-01-19 to 2018-06-23 &  509, 507  \\
  Perkins, Flagstaff       & $B$, $V$, $I$       &  2008-10-23 to 2023-02-10  &  54850 to 59227  &  2009-01-19 to 2021-01-13 & 244, 252, 762 \\
                           & $R$             &  2006-01-06 to 2023-02-12  &  54850 to 59227  &  2009-01-19 to 2021-01-13 &  240  \\
  SMARTS                   & $B$, $V$          &  2008-11-07 to 2017-04-14  &  54854 to 57835  &  2009-01-23 to 2017-03-23 &  534, 533  \\
                           & $R$, $J$          &  2008-02-05 to 2017-04-12  &  54854 to 57835  &  2009-01-23 to 2017-03-23   &  530, 487  \\
                           & $K$             &  2008-04-12 to 2016-03-05  &  55129 to 57452  &  2009-10-25 to 2016-03-05   &  386  \\  
  OAGH, Mexico             & $J$             &  1998-07-06 to 2021-11-15  &  54850 to 59222  &  2009-01-19 to 2021-01-08 &  156  \\
                           & $H$, $Ks$         &  1995-10-22 to 2021-11-15  &  54850 to 59222  &  2009-01-19 to 2021-01-08 &  155, 152  \\
  UMRAO                    & 4.8 GHz       &  2007-11-09 to 2012-06-15  &  54894 to 56045  &  2009-03-04 to 2012-04-28 &  88  \\
                           & 8.0 GHz       &  2007-11-13 to 2012-05-17  &  55132 to 56042  &  2009-10-28 to 2012-04-25 &  115  \\
                           & 14.5 GHz      &  2007-11-18 to 2012-06-24  &  54892 to 55972  &  2009-03-02 to 2012-02-15 &  130  \\
  OVRO                     & 15.0 GHz      &  2008-01-08 to 2023-08-20  &  54850 to 59226  &  2008-04-24 to 2021-01-12 &  529  \\  
  Mets$\ddot{a}$hovi, Finland  & 37 GHz        &  2003-01-03 to 2023-05-16  &  54896 to 59224  &  2009-03-06 to 2021-01-10 &  1375  \\
  VLBA-BU BLAZAR           & 43 GHz        &  2007-06-14 to 2023-06-30  &  54850 to 59222  &  2009-01-19 to 2021-01-08 &  143  \\ 
                           & 86 GHz        &  2020-09-06 to 2022-03-12  &  59222 to 59222  &  2021-01-08 to 2021-01-08 &  1    
    \enddata
\tablecomments{
    Col. (1) The observatory where the data were collected.
    Col. (2) The bands of the data.
    Col. (3) The duration of the data collection, formatted as year, month, and day.
    Col. (4) The duration of the data applied in constructing the SEDs for this work, formatted in $MJD$.
    Col. (5) The duration of the data applied in constructing the SEDs for this work, formatted as year, month, and day.
    Col. (6) The number of data points within the duration specified in either Col. (4) or Col. (5).
  }
\end{deluxetable*}


\section{SED Modeling\label{sec:sed_modelling}}
The observed SED covering the UV to radio bands was modeled with a parabola in the logarithms of the variables (hereafter log-parabola, LP in short). The simplified model assumes that radiation comes from a single region in the jet, filled with chaotic magnetic fields and electrons, moving relativistically at a small angle to the observer's line of sight. 
Note that for blazars, the location of the radio core varies significantly with frequency, particularly across the range of 4.8 to 86 GHz, which we will use in this work. However, this variation is generally smaller for BL Lacs. In the specific case of OJ 287, \cite{Pushkarev_etal_2012} estimate that the 15 GHz core is located within 4.1 pc of the black hole, with the positional difference between the 15 GHz and 8 GHz cores being less than 0.05 mas. Consequently, all cores at frequencies higher than 15 GHz should lie within 4 pc of the black hole. Given this, if the emission region spans about 4 pc, it can reasonably be treated as a single region for modeling purposes.
As a result, the observed radiation experiences Doppler boosting, described by the Doppler factor
 $\delta = [\Gamma (1-\beta\cos{\theta})]^{-1}$, where $\beta$ is the velocity of the source divided by the light velocity, $\Gamma$ is the Lorentz factor and $\theta$ is the angle between the line of sight of observer and direction of motion of the source.
 
 A LP distribution is not only a simple mathematical tool for spectral modeling, but also relates to the physics of the electron acceleration processes. Both the statistical and stochastic acceleration mechanisms can reproduce the electron energy distribution as a LP law, resulting in an LP SED approximately \citep[][and references therein]{2004A&A...413..489M, Tramacere_etal_2007, 2008A&A...478..395M, Tramacere_etal_2011, Chen2014}. The LP function for SED modeling has three spectral parameters and can be defined as 
\begin{equation}
    \log \nu f_{\nu} = -b\ (\log \nu\ -\ \lognup)^2\ +\ \lognupfnup \label{equation_LP} 
\end{equation}
where the $b$ measures the curvature around the SED peak, $\nup$ is the peak frequency, and $\lognupfnup$ is the peak flux \citep{2011MNRAS.417.1881R, Chen2014, Gupta_etal_2016, YangJH_etal_2022}.  
 
The statistical acceleration mechanism framework requires either an energy-dependent acceleration probability ($p_{\rm a}$) or variations in the fractional acceleration gain ($\epsilon$). 
Studies by \citet{2004A&A...413..489M} and \citet{2008A&A...478..395M} demonstrate that a LP spectrum can be obtained when the probability of particle acceleration is energy-dependent. This scenario naturally occurs when particles are confined by a magnetic field whose efficiency decreases as the gyration radii of the particles increase \citep{2011MNRAS.417.1881R}. Additionally, in cases where there are fluctuations in the energy gain parameter $\epsilon$, a LP spectrum can also form under specific conditions if $\epsilon$ is treated as a random variable centered around a systematic value \citep{Tramacere_etal_2011}.

Moreover, a LP spectrum can result from the stochastic acceleration mechanism, described by the Fokker-Planck equation with an included momentum diffusion term \citep{Tramacere_etal_2007, Tramacere_etal_2011}. In this framework, a LP distribution of electron energy can be derived from a `quasi-monoenergetic' injection \citep{Kardashev_1962}.

By maintaining temporal intervals of up to 10 days, we successfully constructed 106 SEDs spanning from UV to radio bands. The choice of a 10-day interval is primarily motivated by the need to balance the quantity of SEDs and the simultaneity of MW data that comprises these SEDs. This choice allows for a 10\% to 18\% increase in the number of SEDs compared to intervals of 4 to 8 days. However, extending the interval beyond 10 days yields less than a 5\% increase in SEDs, while compromising simultaneity across different data filters. Additionally, 10 days correspond to the typical observational window for OJ 287 during a month, especially around the new moon. 

These SEDs cover the $MJD$ range from 54850 (2009-01-19) to 59227 (2021-01-13). The duration of MW data in each band, used for constructing the SEDs in this work, is listed in Cols.~(4) and~(5) of Table~\ref{table:obslog}. The number of data points of each band within the applied durations are also shown in Col. (6), from which we select the data points to construct SEDs. Each SED includes at least one data point in {the following seven series of bands: \\
(i) UV bands ($w2$, $m2$, $w1$, $u$), \\
(ii) $B$, \\
(iii) $V$, \\
(iv) $R$, \\
(v) partial optical plus partial NIR bands ($I$, $J$, $H$), \\
(vi) additional NIR bands ($K$, $Ks$), and \\
(vii) radio bands (86 GHz, 43 GHz, 37 GHz).

For each band within each series of bands, if multiple measurements are available from one observatory or different observatories, the final flux for that band is calculated as the median of these measurements. Among the 106 SEDs, 72 SEDs includes at least one data point in additional radio bands (15.0 GHz, 14.5 GHz, 8.0 GHz, 4.8 GHz). Galactic extinction correction was performed for the data in the NIR to UV bands \citep{Cardelli89, Schlegel98}, and redshift correction was subsequently performed for the constructed SEDs.

We fit the SEDs using the LP model with the maximum likelihood method, which minimizes the negative log-likelihood. This is implemented using the {\tt optimize.minimize} function \citep{Virtanen_etal_2020}. The negative log likelihood function is 
\begin{equation}
- \log \mathcal{L} = -\frac{1}{2} \sum \left( \frac{(y - y_{\text{model}})^2}{\sigma^2} + \log(\sigma^2) \right)
\end{equation}
\begin{equation}
\sigma^2 = y_{\text{err}}^2 + (y_{\text{model}} \exp(\text{log\_f}))^2, 
\end{equation}
where $y$ means the observed $\log\nu f_{\nu}$ values, $y_{\rm model}$ means the predicted $y$ value obtained from the model shown in Eq.~\ref{equation_LP}, and $\logf$ is a parameter representing an additional scatter beyond the measurement error $y_{\rm err}$. The total uncertainty $\sigma^2$, is calculated as the square of the measurement uncertainty $y_{\rm err}^2$, plus an additional term that scales with the model value and the exponential of $\logf$. With $\logf > 0$, the model uncertainty allows for more flexibility to account for additional scatter not captured by $y_{\rm err}$. In practice, incorporating $\logf$ during the fitting helps balance between under-fitting by providing too little flexibility and over-fitting in the model.

The significance of our SED model fitting is first evaluated using the reduced $\chi^2$, the sum of squared normalized residuals divided by the degrees of freedom. In our case, the reduced $\chi^2$ are greater than 1, indicating larger residuals than expected from the uncertainties. For further investigation, we used the probability based on $\logf$ as a model-adjusted flexibility measure to account for additional scatter. The $\logf$ values, ranging from 0 to 0.013 with a median of 0.005, result in a small increase in the total uncertainties $\sigma^2$ of up to 2.6\% with a median of 1.0\%. These low $\logf$ values confirm that the larger reduced $\chi^2$ values are likely caused by slightly underestimated measurement uncertainties. Overall, the fitting remains generally significant.

A Monte Carlo approach is applied to estimate the uncertainties of the fitting parameters. For each SED, 50 random mock SEDs are generated by introducing Gaussian noise to the original SED. At each frequency $\log \nu$, in a given mock SED, the noise term is randomly drawn from a normal distribution with the observed $\log \nu f_{\nu}$ error as the standard deviation. We then fit each mock SED using the same fitting strategy. The 1$\sigma$ dispersion of the measurements relative to the original values is taken as the corresponding uncertainty.
Together with visual check, we define SEDs with $b > 0.02$ as well-fit by the LP function. All the 106 SEDs can be well fit by the LP model, as shown in Fig.~1 and Fig.~2.  

Between $MJD$ 55558 and 55621, although the $K$-band data points of 6 SEDs deviate most significantly from the model fits, our analysis shows that their inclusion does not significantly impact the overall results, as comparisons of fits with and without these data points reveal little difference in the derived parameters. The relative bump in the $K_{s}$ band is generally attributed to thermal emission from dust at a wide range of temperatures \citep{2004ASPC..311...37W}, contribution of the host galaxy of the blazar, and IR contribution of torus \citep{2024ApJ...963...48G}.

With the number of data points in radio bands of each SED shown in the upper left corner of each panel in Fig.~\ref{fig:sed_class1_0} and Fig.~\ref{fig:sed_class1_1}, we investigated whether the obtained $b$ values are influenced by the number of data points in radio bands. We find that only 12 epochs have a single data point in radio. When all the 106 SEDs are ranked in order of decreasing $b$, none of the 30 highest epochs have only a single radio data point, but 7 of the lowest epochs do. To further check the influence of the number of radio data points, as 105 SEDs except the 1st SED include data points at 37 GHz, we reduced the number of radio data points at all epochs to one, i.e., the data point at 37 GHz is selected if available, otherwise the closest data point in frequency as 43 GHz is chosen. This simplification resulted in updated $b$ values ranging from 0.034 to 0.210, with a median of \( 0.104 \pm 0.043 \), as shown by the green histogram in Fig.~\ref{fig:b_b37G}. For comparison, the original $b$ values ranged from 0.038 to 0.234, with a median of \( 0.117 \pm 0.045 \), shown by the grey histogram. The difference between the median values is \( -0.013 \), smaller than the standard deviation around the median. The KS test yielded a test statistic of 0.18 and a p-value of 0.06, indicating that at the 5\% significance level, there is no statistically significant difference between the two distributions.


\begin{figure*}
\hspace{-1cm}
 \centering
 \includegraphics[scale=0.5]{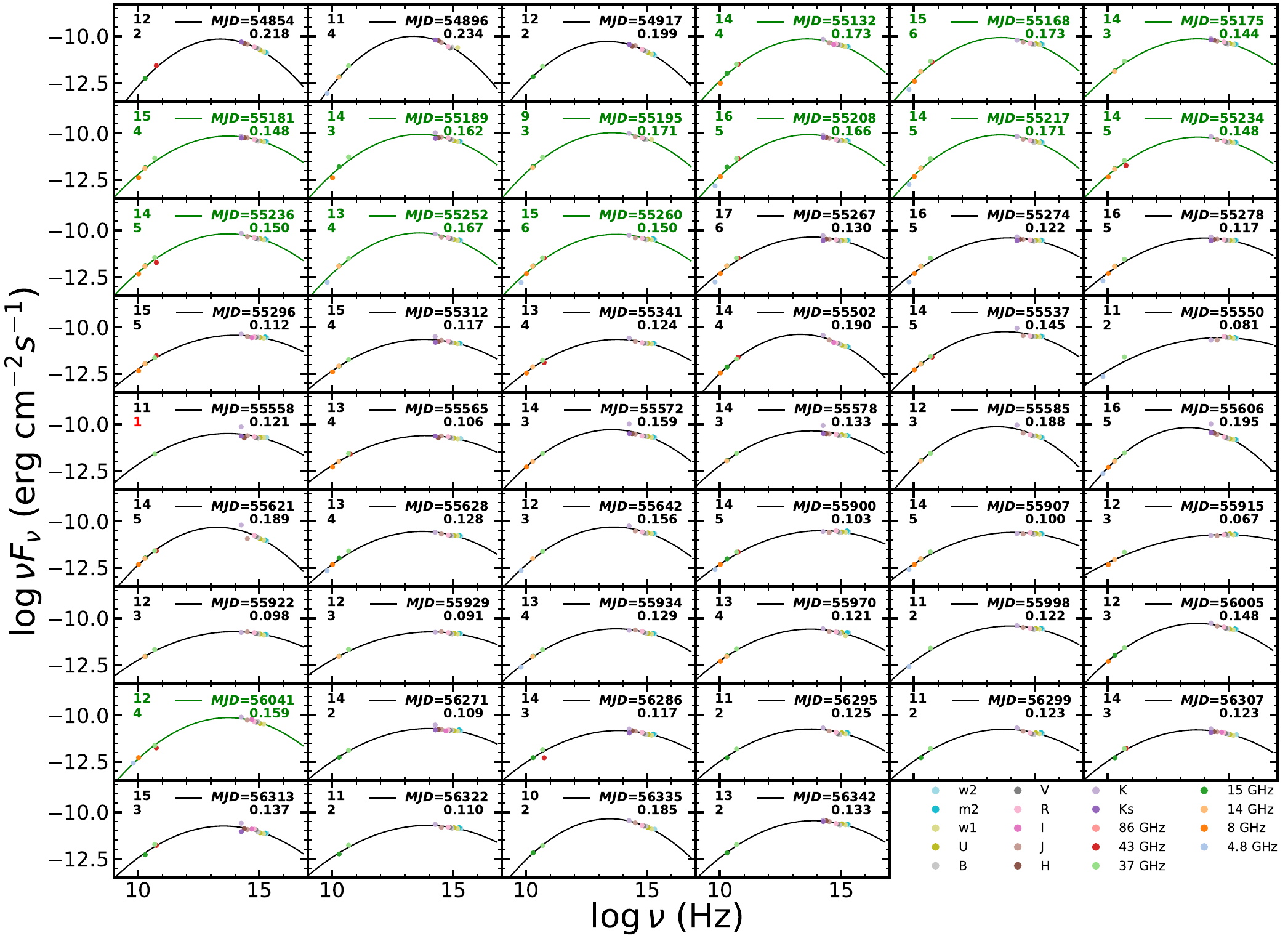}
    \caption{52 SEDs which can be well fit by LP model with the spectral curvature $b$ larger than 0.02. The centered $MJD$ values and the corresponding ($b$) values are indicated in the upper-right corner of each panel. The data points for each SED were observed within a time range of the listed $MJD$ $\pm$ 5 days. The texts in the upper-left corner of each panel indicate the total number of data points and the number of radio data points included in each SED.} The modeled SEDs shown in green correspond to those located within the flare segments.
    \label{fig:sed_class1_0}
\end{figure*}

\begin{figure*}
\hspace{-1cm}
 \centering
 \includegraphics[scale=0.5]{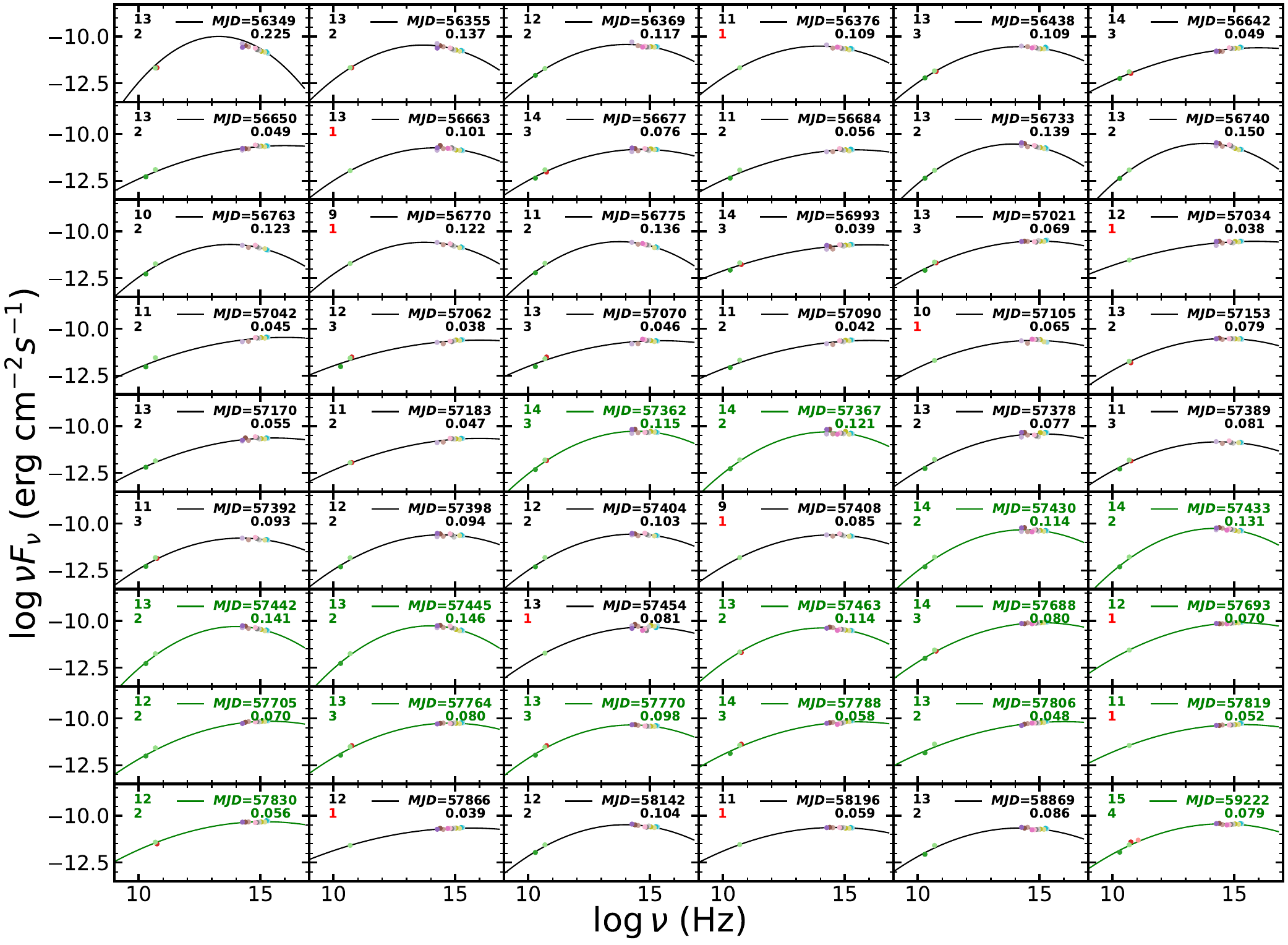}
    \caption{Same with Figure~\ref{fig:sed_class1_0}, additional 54 SEDs which can be well fit by LP model with $b$ greater than 0.02.} 
    \label{fig:sed_class1_1}
\end{figure*}

\begin{figure}
    \begin{minipage}{0.48\textwidth}
        \includegraphics[width=\linewidth]{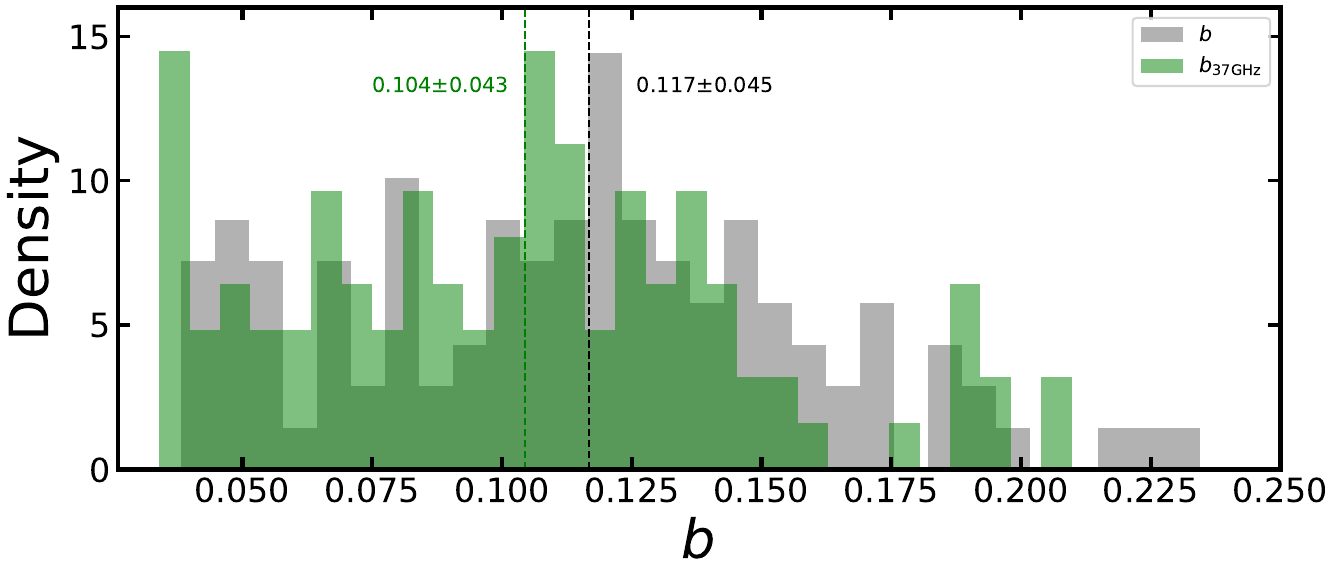}
        \caption{Distributions of $b$ and $b_{\rm 37G}$ are represented by the grey and green histograms, respectively. The $b$ values are derived from the LP model fitting applied to all the constructed SEDs. In contrast, the $b_{\rm 37G}$ values are estimated by fitting the LP model to SEDs where the number of radio data points is reduced to one. Specifically, the data point at 37 GHz is selected if available; otherwise, the closest available frequency, such as 43 GHz, is used.}
        \label{fig:b_b37G}
    \end{minipage}%
\end{figure}
\section{Results\label{sec:results}}
\subsection{Flare and Quiescent States\label{subsec:states}}
As shown in the upper panel of Fig.~\ref{fig:vmags_mjd}, the target OJ 287 shows optical variability in $V$ band across the $MJD$ range of 54000 to 60000. For almost three months each year, OJ 287 is not visibly accessible to the optical telescopes used to collect data for this study. By analyzing the $V$-band flux distribution of OJ 287, we identified a distinct log-normal profile, as shown in Fig.~\ref{fig:flux_distr}. We first determined the optimal number of Gaussian components using the Bayesian Information Criterion, which indicated that a single Gaussian component was most appropriate. We then fit a Gaussian Mixture Model using this optimal number of components and extracted the mean and standard deviation ($\sigma$) from the fitted profile. Using these parameters, we established a flux limit based on the mean plus half the $\sigma$ of the distribution, resulting in a value of $10^{-10.49}\ {\rm erg\ cm^{-2}\ s^{-1}}$ (shown as the right edge of the green region in Fig.~\ref{fig:flux_distr} and also the horizontal green line in the upper panel of Fig.~\ref{fig:vmags_mjd}). The cumulative distribution function at the flux limit is 0.69, indicating the probability that a randomly selected sample will have a value less than or equal to the flux limit. 

We defined a `flare' segment in the $V$-band light curve as any observation period containing more than three consecutive data points with flux exceeding a specified limit, with segments not meeting this criterion designated as `quiescent'. Testing variations from one to six consecutive data points revealed that the number of flare segments fluctuates only slightly, by 2 to 4 segments, without affecting the number of SEDs within the flare segments or the duration proportion of the flare segments. This indicates that the choice of consecutive data points does not influence the subsequent analysis of SEDs in flare versus quiescent segments. Our choice of three consecutive data points strikes a balance by minimizing misclassification from isolated outliers, ensuring genuine flare detection, and maintaining enough segments for meaningful analysis, making it an optimal threshold. The flare segments in the $V$ band are shaded in green in Fig.~\ref{fig:vmags_mjd}; the start and stop dates of the individual flare segments are listed in Table~\ref{table:flarestate}.

This categorization results in 19 flare segments with durations ranging from 1 to 312 days. As shown in Fig.~\ref{fig:vmags_mjd}, there are no constructed SEDs available for 10 out of the 19 identified flare segments, including the first and longest flare segment with a duration of 312 days. This absence is partly due to the scarcity of data points in the UV to radio bands.

The flare segments summarized in Table~\ref{table:flarestate} differ from those discussed in the introduction, which are used for orbit determination. Only two of these segments, SEDs with central $MJD$ of 57362 and 57367 (Dec. 6 and Dec. 11, 2015), coincide with the time range of the predicted and confirmed flare in 2015, as shown in the fifth row, third to fourth column of Fig.~\ref{fig:sed_class1_1}, and listed as the 13th row in Table~\ref{table:flarestate}. Both flares are exceptionally bright and exhibit rapid variability, requiring higher temporal resolution for studying their spectral changes \citep{2016ApJ...819L..37V}. Although they could have been excluded from the adopted flare segments, their inclusion as two single epochs does not affect the results of this study.


\begin{figure}
    \begin{minipage}{0.48\textwidth}
        \includegraphics[width=\linewidth]{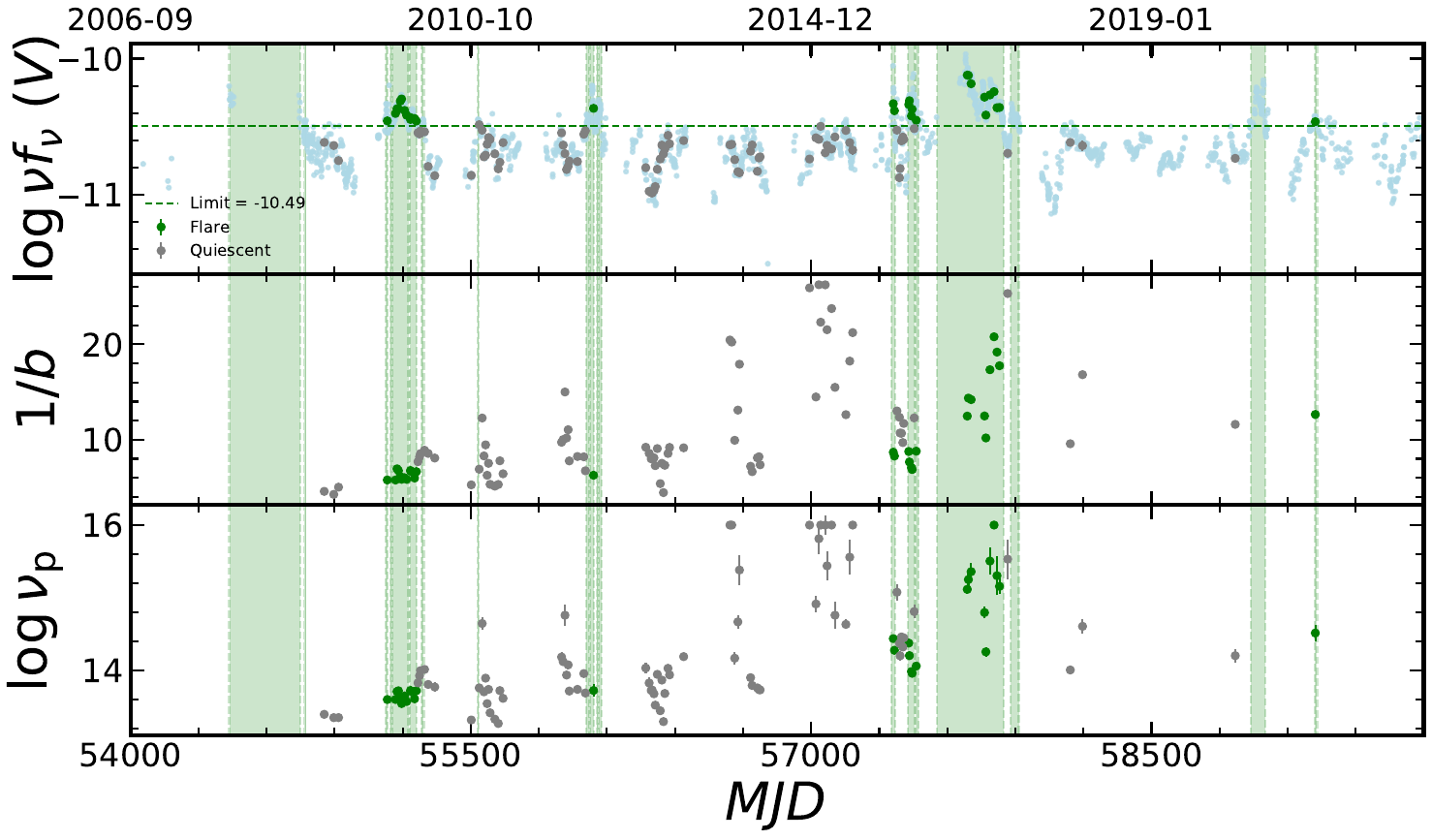}
        \caption{Upper: $\log \nu f_{\nu} (V)$ v.s. $MJD$, The green regions refer to the time ranges defined as flare segments. The light blue dots refer to the values of the constructed SEDs, while the green and grey dots refer to those of the SEDs in the flare and quiescent segments, respectively. Middle: $1/b$ v.s. $MJD$. Lower: $\lognup$ v.s. $MJD$. Both $1/b$ and $\lognup$ vary with $MJD$, exhibiting similar patterns.}
        \label{fig:vmags_mjd}
    \end{minipage}%
\end{figure}

\begin{figure}
    \begin{minipage}{0.48\textwidth}
        \includegraphics[width=\linewidth]{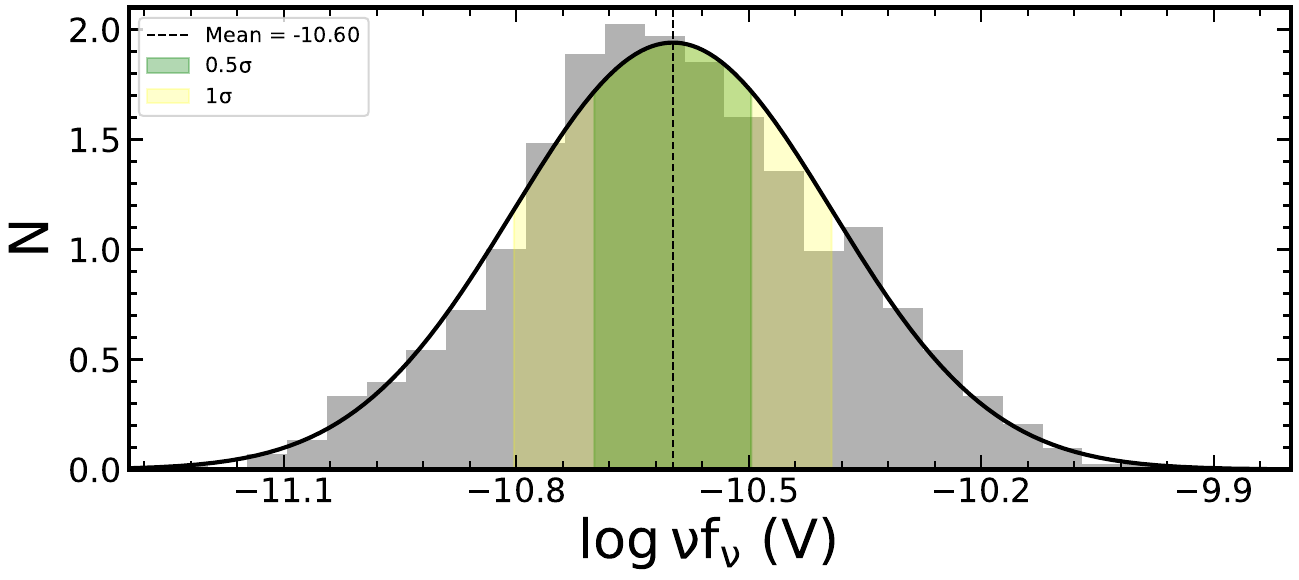}
        \caption{$V$-band flux distribution of OJ 287. The shaded grey region represents the observed distribution, and the black curve is the fitted log-normal model. The vertical dashed line indicates the mean of the model. The green and yellow regions correspond to the 0.5$\sigma$ and 1$\sigma$ ranges around the mean, respectively.}
        \label{fig:flux_distr}
    \end{minipage}%
\end{figure}

The total duration of the flare segments spans 954 days, representing 15\% of the 6475-day time span of the $V$-band light curve analyzed in this work. Among the 106 SEDs constructed from nearly simultaneous multi-band photometric data, 30 SEDs occur during flare segments (modeled SEDs in green in Fig.~\ref{fig:sed_class1_0} and Fig.~\ref{fig:sed_class1_1}), while 76 SEDs are in quiescent states (modeled SEDs in grey). 

\subsection{SEDs at Different States\label{subsec:seds_separate}}
We compared the SEDs in flare and quiescent segments based on three key parameters: the peak intensity ($\lognupfnup$), the SED curvature ($b$), and the peak frequency ($\lognup$). As shown in Fig.~\ref{fig:compare_flare_quiescent}, the median $\lognupfnup$ for flare segments is $0.37 \pm 0.22$ dex higher than for quiescent segments. The median curvature $b$ is slightly larger in flare segments ($0.14$) compared to quiescent segments ($0.11$). However, this difference is negligible when considering the uncertainty of the median value ($\sim 0.04$). Similarly, $\lognup$ values remain consistent, with $14.02 \pm 0.70$ for flare segments and $13.95 \pm 0.79$ for quiescent segments, respectively. Here, the uncertainty of the median value are derived from the one $\sigma$ dispersion of the distribution of the corresponding parameter.

\begin{figure}
    \begin{minipage}{0.48\textwidth}
        \includegraphics[width=\linewidth]{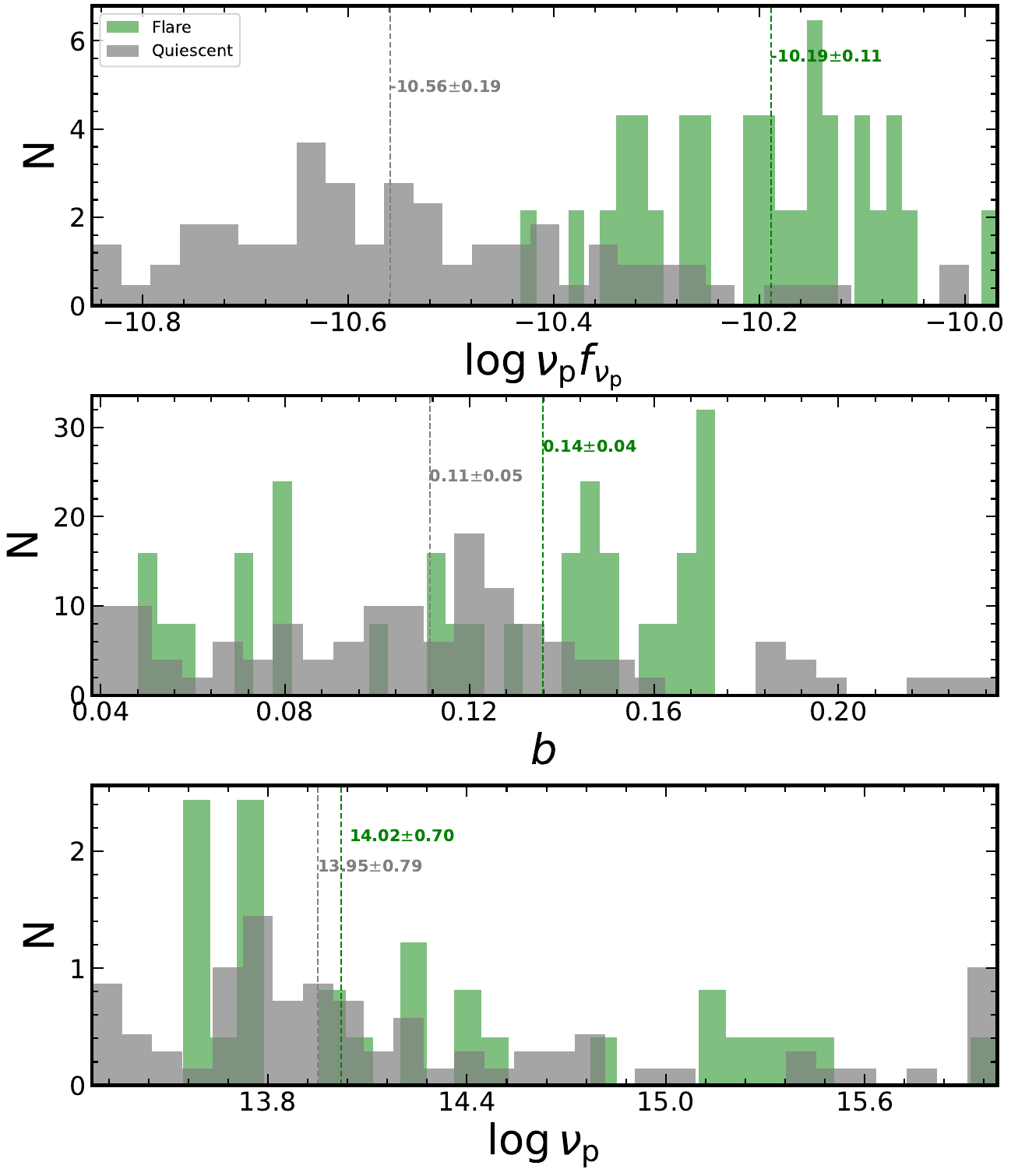}
        \caption{Upper: Distributions of $\lognupfnup$ for SEDs in flare (green) and quiescent (grey) segments. Middle: Distributions of curvature $b$ for SEDs. Lower: Distributions of $\lognup$ for SEDs. The median $\lognupfnup$ for SEDs in flare segments is larger than that for SEDs in quiescent segments by $0.37 \pm 0.22$ dex, while $b$ and $\lognup$ are consistent within their uncertainties.}
        \label{fig:compare_flare_quiescent}
    \end{minipage}%
\end{figure} 

\begin{deluxetable}{ccccccc}
  \tabletypesize{\footnotesize}
  \tablecaption{Start and Stop Dates of the individual Flare Segments \label{table:flarestate}}
  \tablewidth{0pt}
  \tablehead{\colhead{ } & \colhead{Start Date} & \colhead{Stop Date}  & \colhead{Start $MJD$} & \colhead{Stop $MJD$} & \colhead{$N_{\rm data}$} & \colhead{$N_{\rm SED}$} \\
  \colhead{(1)} & \colhead{(2)} & \colhead{(3)} & \colhead{(4)} & \colhead{(5)} & \colhead{(6)} & \colhead{(7)} }  
\startdata
    1 & 2007-12-01 & 2008-10-08 & 54435 & 54747 & 14 & 0 \\
    2 & 2008-10-29 & 2008-10-30 & 54768 & 54769 & 4 & 0  \\
    3 & 2009-10-21 & 2009-10-29 & 55125 & 55133 & 17 & 1 \\
    4 & 2009-11-12 & 2009-11-15 & 55147 & 55150 & 5  & 0 \\
    5 &  2009-11-18 & 2010-01-29 & 55153 & 55225 & 76 & 7 \\
    6 &  2010-02-04 & 2010-03-07 & 55231 & 55262 & 25 & 4 \\
    7 &  2010-03-28 & 2010-04-07 & 55283 & 55293 & 9  & 0 \\
    8 &  2010-12-01 & 2010-12-06 & 55531 & 55536 & 4  & 0 \\
    9 &  2012-03-24 & 2012-04-05 & 56010 & 56022 & 9  & 0 \\
    10 &  2012-04-09 & 2012-04-27 & 56026 & 56044 & 24 & 1 \\
    11 &  2012-05-08 & 2012-05-13 & 56055 & 56060 & 6  & 0 \\
    12 &  2012-05-15 & 2012-05-28 & 56062 & 56075 & 19 & 0 \\
    13 &  2015-11-27 & 2015-12-13 & 57353 & 57369 & 17 & 2 \\
    14 &  2016-02-09 & 2016-03-09 & 57427 & 57456 & 33 & 4 \\
    15 &  2016-03-13 & 2016-03-25 & 57460 & 57472 & 15 & 1 \\
    16 &  2016-06-14 & 2017-04-04 & 57553 & 57847 & 230& 9 \\
    17 &  2017-05-05 & 2017-06-11 & 57878 & 57915 & 30 & 0 \\
    18 &  2020-03-28 & 2020-05-31 & 58936 & 59000 & 46 & 0 \\
    19 & 2021-01-04 & 2021-01-16 & 59218 & 59230 & 12 & 1 \\
\enddata
    \tablecomments{
      Col.(1) Indices of the flare segments.
      Col. (2-3) Start and stop date of each flare segment.
      Col. (4-5) Start and stop $MJD$ of each flare segment.
      Col. (6) Number of data points in the $V$-band light curve within each flare segment.
      Col. (7) Number of constructed SEDs within each flare segments.}
\end{deluxetable}

\subsection{Color Variability\label{subsubsed:color-varia}}
As variations in the optical flux of blazars are accompanied by spectral changes, studying the color index--magnitude (CM) relation can help to understand the origin of the variability in blazars. Earlier studies have found significant bluer-when-brighter (BWB) / redder-when-brighter (RWB), and achromatic trends on diverse time-scales CM diagram \citep[e.g.][and references therein]{2006A&A...450...39G,Gaur_etal_2012, Agarwal_etal_2016,2019MNRAS.488.4093A,2021A&A...645A.137A}.

Due to the potential for non-negligible magnitude fluctuations when switching filters during non-simultaneous observations, making accurate color measurements difficult, it is necessary to obtain very dense and precise simultaneous multiband observations to detect weak CM relationships. Based on the $V$ band magnitudes and $B-V$ color indices of the 106 SEDs, as shown in Fig.~\ref{fig:color_V}, we found there is a weak BWB relation, with a Spearman correlation coefficient $r \sim$ 0.28 at a confidence level above 99.6\% for the SEDs. Further excluding the three outlier points with significant $B-V$ difference and constraining $0 < B-V < 0.6$, we achieved $r \sim 0.26$ with a confidence level exceeding 99.2\%. This is also confirmed by a weak anti-correlation between the SED peak frequency $\lognup$ and the $V$ band magnitude (Fig.~\ref{fig:vmag_nupeak}), showing the peak frequency being higher at brighter magnitude, i.e., $r \sim -0.19$ at a confidence level above 94.6\%. 

Optical emission from blazars typically consists of contributions from both the relativistic jet and the accretion disk, with the jet often being the dominant. When synchrotron radiation from the relativistically boosted jet outshines the emission from the disk, the BWB trend can be attributed to either the acceleration of relativistic particles or the injection of fresh electrons with an even harder energy distribution \citep{Kirk_etal_1998, Mastichiadis_etal_2002, Fiorucci_etal_2004, Gupta_etal_2016}. For the RWB trend, the contribution of the accretion disk to the total emission could be significant. In general, BWB and RWB trends were found in BL Lacs  and FSRQs, respectively \citep[e.g.][and references therein]{Gaur_etal_2012,2019MNRAS.488.4093A,2021A&A...645A.137A}, but sometimes opposite trend is also noticed \citep[e.g.][and references therein]{Gaur_etal_2012}. 

In Fig.~\ref{fig:color_V}, we found a stronger BWB trend during flares (green symbols) compared to quiescent states (grey symbols), indicating a dominance of jet over the accretion disk in the flare segments. In flare and quiescent segments, the correlation coefficient is $r \sim 0.40$ at a confidence level over 97.3\% and $r \sim 0.20$ at a confidence level above 91.1\%, respectively. This pattern is corroborated by anti-correlations between $\log\nu_\mathrm{peak}$ and the $V$-band magnitude (Fig.~\ref{fig:vmag_nupeak}), with $r \sim -0.53$ and a confidence level above 99.7\% in flares, compared to $r \sim -0.30$ and a confidence level over 99.2\% in quiescent segments.

The Doppler factor variations are also usually attributed to the achromatic behavior, and this interpretation is most likely supported by the geometric scenario \citep[e.g.][]{2002A&A...390..407V}. \citet{2021A&A...654A.169L} estimated Doppler factor versus frequency in log-log space for 61 blazars, including OJ 287. They used five radio band data from 4.8 GHz to 37 GHz and found there is a linear relation with slope 0.22$^{+0.29}_{-0.29}$, intercept 1.07$^{+0.32}_{-0.35}$ and a Pearson correlation coefficient 0.58.
This linear relation may be extended from NIR to UV bands to estimate the Doppler factor in these EM bands.

\begin{figure}
    \begin{minipage}{0.48\textwidth}
        \includegraphics[width=\linewidth]{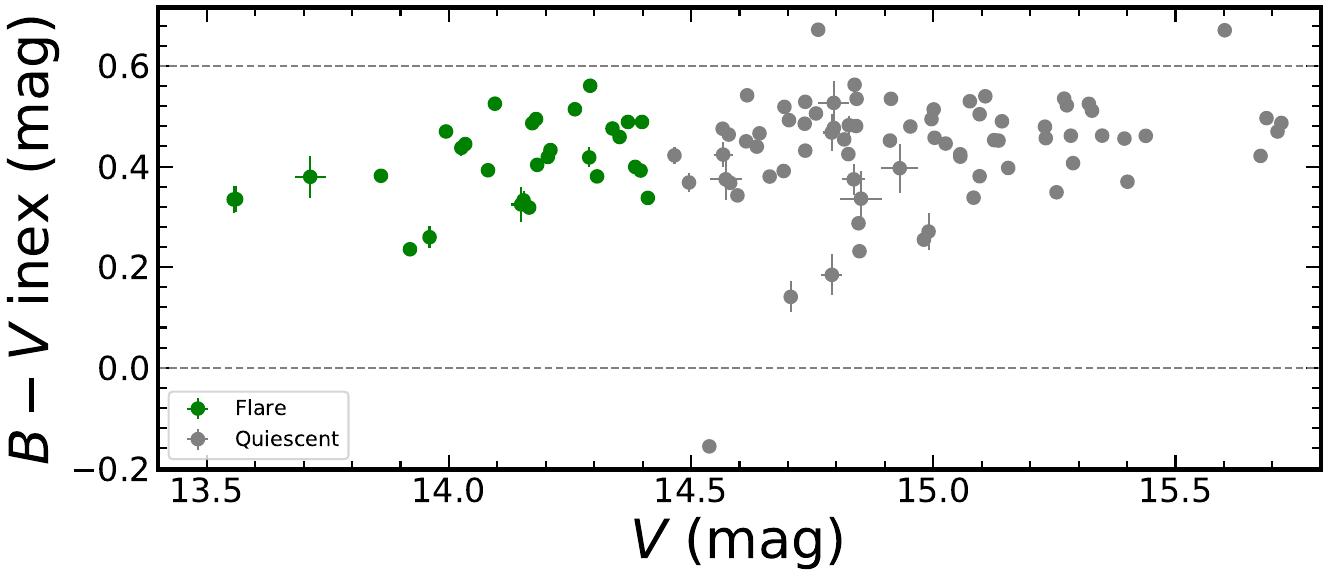}
        \caption{$B-V$ index v.s. $V$ band magnitude. There is a bluer-when-brighter relation, with $r \sim$ 0.28 at a confidence level above 99.6\%. Further excluding the three outlier points with significant $B-V$ difference by constraining $0 < B-V < 0.6$ (data points between the two dashed horizontal lines), we achieved $r \sim 0.26$ with a confidence level exceeding 99.2\%. The BWB trend is stronger during flare segments (green symbols) compared to quiescent ones (grey symbols), with $r\sim0.40$ at a confidence level over 97.3\% versus $r\sim0.20$ at a confidence level over 91.1\%.}
        \label{fig:color_V}
    \end{minipage}%
\end{figure}

\begin{figure}
    \begin{minipage}{0.48\textwidth}
        \includegraphics[width=\linewidth]{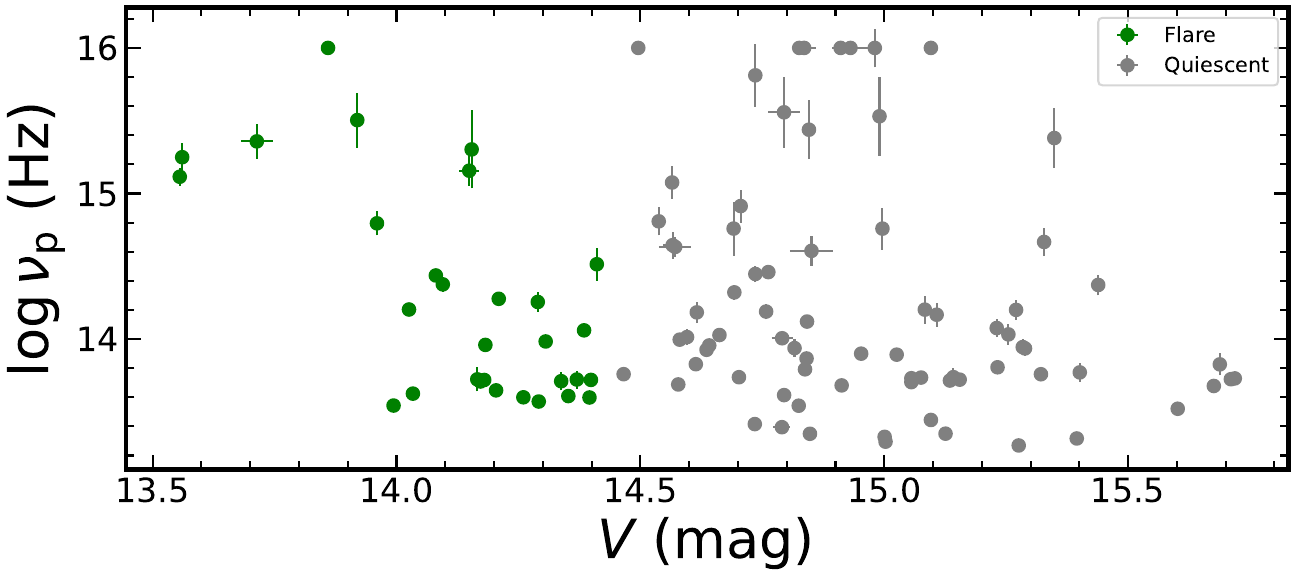}
        \caption{$\lognup$ vs. $V$ band magnitude, where $r$ is -0.19 at a confidence level above 94.6\%. The trend is stronger during flare segments (green symbols) compared to quiescent ones (grey symbols), with $r\sim-0.53$ at a confidence level over 99.7\% versus $r\sim-0.30$ at a confidence level over 99.2\%. } 
        \label{fig:vmag_nupeak}
    \end{minipage}%
\end{figure}

\section{Discussion\label{sec:discussion}}
\subsection{LP SEDs and Statistical Particle Acceleration\label{subsec:discussion-model}}
The study of LP SEDs in blazars has uncovered significant correlations. For a sample of 60 blazars, radio to X-ray SEDs were well fitted by the LP model, where the peak frequency was found to be anti-correlated with bolometric luminosity \citep{1996ApJ...463..444S}. In contrast, for a sample of 300 BL Lacs, SEDs from radio to X-ray also fit the LP model, showing an anti-correlation between the peak frequency and flux at radio (5 GHz) and optical (5500 \AA), but no such anti-correlation was observed with X-ray flux \citep{2006A&A...445..441N}.

Additional studies have also explored the connection between peak frequency and curvature. By fitting the SED from radio to optical with a LP model for a sample of 18 blazars, \cite{Landau_etal_1986} found an anti-correlation between peak frequency and curvature for the 15 blazars which can be well fit by a LP model. Similar results were obtained by recent works \citep{2011MNRAS.417.1881R, Chen2014, YangJH_etal_2022}. 

Mainly there are two different scenarios explaining the correlation between the peak frequency and curvature. The first scenario is within the framework of statistical acceleration. For the case of the energy-dependent acceleration probability ($p_{\rm a}$), \citet{2004A&A...413..489M} showed that if $p_{\rm a}$ is inversely related to the particle's energy, the resulting SED naturally adopts a LP form, where the curvature $b$ can be inversely correlated with the peak energy or frequency ($\lognup$), described by $1/b \propto 5/2\ \lognup $. In contrast, considering fluctuations in the fractional energy gain ($\epsilon$), \citet{Tramacere_etal_2011} demonstrated that treating $\epsilon$ as a random variable around a systematic energy gain also leads to an inverse relationship between $b$ and $\lognup$, following the relation $1/b \propto 10/3\ \lognup$.

The second scenario is within the framework of stochastic acceleration mechanism, which can predict an anti-correlation between $b$ and $\log \nu_{\rm p}$, described by the relation $1/b \propto 2 \log \nu_{\rm p} $ \citep{Tramacere_etal_2007, Tramacere_etal_2011}. 

Using a sample of 10 low- to intermediate-synchrotron-peaked blazars, \cite{2011MNRAS.417.1881R} found an anti-correlation between $b$ and $\nu_{\rm p}$, and suggested that the log-parabolic (LP) SED shape is likely characterized by a full statistical acceleration mechanism acting on the emitting electrons. While using a large sample of 48 blazars, \cite{Chen2014} found that the slope of the correlation between $1/b$ and $\nu_{\rm p}$ as $2.04 \pm 0.03$ is consistent with the prediction of the stochastic acceleration scenario ($\sim 2$). This is further confirmed by \cite{Anjum_etal_2020}, which found that BL Lacs show a strong signature of stochastic acceleration compared to FSRQs.

In Fig.~\ref{fig:b_a_b_nupeak}, we observe a strong anti-correlation between $b$ and $\lognup$, with $r=-0.95$ at a confidence level above 99.9\%, as shown by the green and grey points. This tendency is also evident in the lower two panels of Fig.~\ref{fig:vmags_mjd}, where both $1/b$ and $\lognup$ vary with the $MJD$ values in similar patterns.
By performing a linear fit to $1/b$ and $\lognup$, we derive the relation  $1/b = (6.20\pm0.08)\lognup - (77.82\pm1.03$) (represented by the solid line). If we exclude the seven data points in the upper right corner with significant $1/b$ differences and constrain $1/b<22$, the slope decreases to $5.79 \pm 0.06$ (represented by the dashed line). This revised slope aligns more closely with the predicted value of $10/3$ from the statistical acceleration mechanism, which accounts for fluctuations in the fractional acceleration gain $\epsilon$, though a notable discrepancy remains, suggesting the influence of additional factors.

The observed slope is significantly steeper than both the previously reported value of 2.04 by \citet{Chen2014} and the theoretical prediction of 10/3 \citep{Tramacere_etal_2011}. Observationally, \citet{Chen2014} derived their slope using less simultaneous data spanning a broad wavelength range (radio to gamma rays) from 48 blazars, including both BL Lacs and FSRQs. While our study focuses on the single blazar OJ 287, with nearly simultaneous data but limited to a narrower range (radio to UV bands). Moreover, the steeper slope in our results may stem from uncertainties in estimating the SED peak frequency and curvature, due to sparse data between the radio and UV bands. A much closer estimate of the observed slope to the theoretical one may be achieved with a large number of blazars' SEDs with much denser data coverage in frequency and time in future.

Nevertheless, the discrepancy with the theoretical predictions may reflect additional physical factors beyond the standard statistical acceleration mechanism, which primarily considers electron acceleration processes in the jet. For example, deviations from idealized conditions or radiative contributions, such as thermal emission from the accretion disk in the optical-UV bands, especially during quiescent states, could contribute to the observed steeper slope. These results highlight the importance of incorporating additional complexities and/or exploring alternative explanations, rather than strictly adhering to the standard statistical acceleration mechanism.

\begin{figure}
    \begin{minipage}{0.48\textwidth}
        \includegraphics[width=\linewidth]{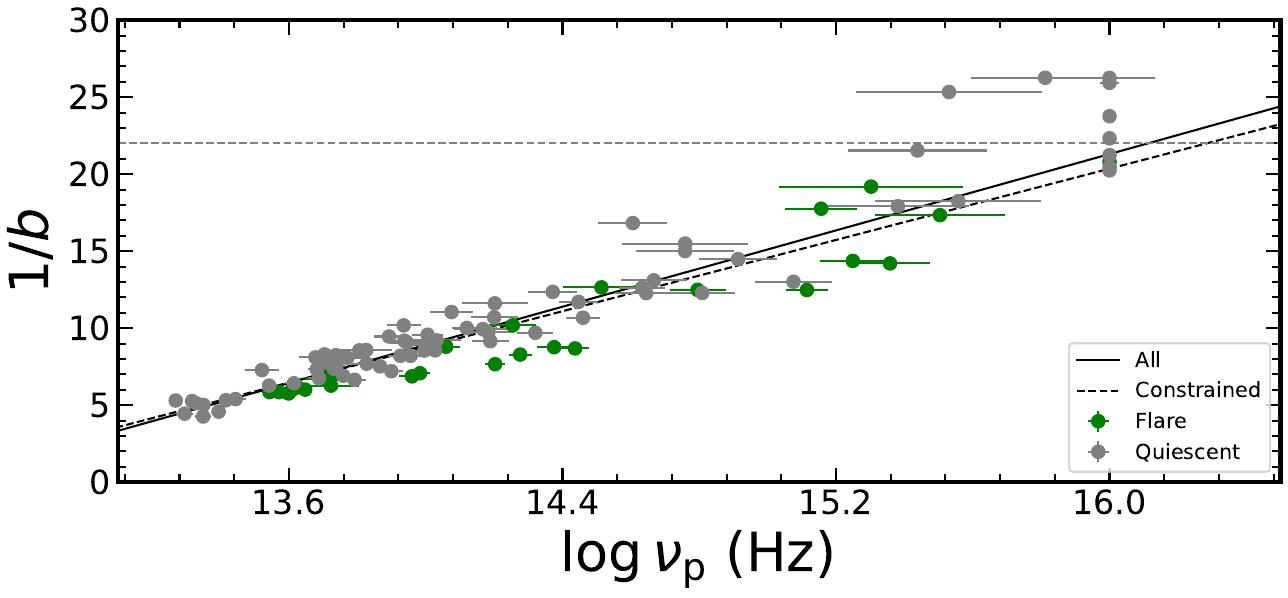}
        \caption{Curvature ($b$) vs. peak frequency ($\log \nu_{\rm p}$). A strong
        anti-correlation is found in the SEDs well fit by the LP model, with a Spearman coefficient of $r = -0.95$ at a confidence level above 99.9\%. The linear fit slope is $6.20 \pm 0.06$, as shown by the solid line. When excluding seven data points in the upper right, where $1/b > 22$, the slope decreases to $5.79 \pm 0.07$, as shown by the dashed line.} 
        \label{fig:b_a_b_nupeak}
    \end{minipage}%
\end{figure}

\subsection{The Cause of SED Changes\label{subsec:discussion-SEDchanges}}
Under the frame of the statistical acceleration mechanism, there are several possible reasons which can explain the changes in the low-energy-peak SEDs along with the time.  
If the changes in the SEDs are primarily caused by a gradual change in the electron energy density distribution due to the synchrotron and IC losses, with no other injections during the period, one would expect a positive relation between the peak intensity and the peak frequency changes \citep{2011MNRAS.417.1881R}. Adopting the epoch where OJ 287 is faintest in the $V$ band among all the considered SEDs as the reference epoch, we calculate the differences in the peak intensity ($\Delta \lognupfnup$) and the peak frequency ($\Delta \lognup$) relative to the reference epoch.
We find that there is a significant anti-correlation between $\Delta \lognupfnup$ and $\Delta \lognup$, i.e., the Spearman correlation coefficient $r \sim -0.38$ at a confidence level above 99.9\%. The discrepancy from the prediction suggests that the evolution of electron energy or electron injection may not be the primary driver of the SED changes. Note that the anti-correlation shows a hint that OJ 287 follows the blazar sequence,  the anti-correlation between the $\lognup$ and $\lognupfnup$ for a blazar sample \citep{Fossati_1998} and related to the physical conditions in the jet \citep{Ghisellini_1998}.

Moreover, the change of other parameters, including the Doppler boosting factor or the magnetic field, may also cause the change in SED along with time. By studying the correlation between the change in peak intensity and the change in both the Doppler boosting factor and the magnetic field for a sample of ten blazars, \cite{2011MNRAS.417.1881R} conclude that the change in the Doppler factor is a strong driver of SED changes, whereas changes in the magnetic field strength may influence only BL Lacs but not all blazars. 

It is found that the Doppler factor is substantially higher in the flaring states of blazars, which may cause the strong increase in the Compton dominance as the external photon density in the co-moving frame of the jet depends on the Doppler boosting factor \citep{2021MNRAS.504.5074S}. Either a fresh injection or re-acceleration of the faster-moving emitting region during propagation could cause it to emit near the central source. Geometrical effects, such as when the jet zones have various orientations, such as in the case of jets in a jet model \citep{2009MNRAS.395L..29G} or twisted in-homogeneous jet model \citep{2017Natur.552..374R}, could also be the cause of the Doppler boosting factor rise. Therefore, the Doppler boosting factor is increased because during the flares, photons may be emitted in a zone observed at smaller angles than the entire jet. 

The effects of magnetic field topology on the SEDs in blazars demonstrate that, in the case of a purely oblique field, the synchrotron component is annulated if the magnetic field is aligned along the line of sight (in the plasma frame) \citep{2020ApJ...898...11J}. However, the impact of an oblique field is diminished and the same effect is not noticed in the presence of a disordered component \citep{2020ApJ...898...11J}. 

In the case of the BL Lac OJ 287, we found that the median $\lognupfnup$ for flare segments is 0.37$\pm$0.22 dex higher than that for quiescent segments, while the median $\lognup$ and $b$ remain consistent within their uncertainties, suggesting that flaring might be mainly caused by the Doppler factor and the ambient magnetic field. This also explains the observed stronger BWB trend in the flare segments compared to the quiescent ones. Therefore, we argue that the possible contribution from variations in the Doppler factor and the magnetic field strength to the observed SED changes. However, we are not able to quantify their contributions based solely on the SED changes. 

\begin{figure}
    \begin{minipage}{0.48\textwidth}
        \includegraphics[width=\linewidth]{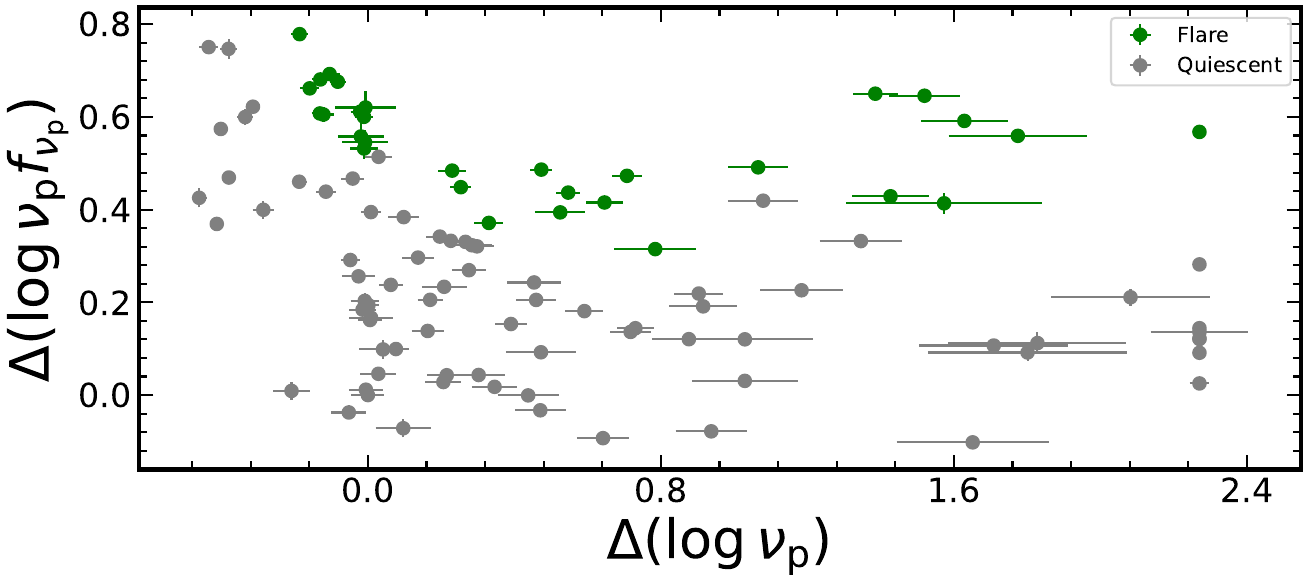}
        \caption{Change in $\lognupfnup$ vs. change in $\lognup$ relative to those in the reference epoch, when OJ 287 is faintest in $V$ band among all the constructed SEDs. The green and grey symbols refer to values of the SEDs in the flare and quiescent segments, respectively. The Spearman correlation coefficient $r$ is -0.38 at a confidence level above 99.6\%. } 
        \label{fig:deltanu_deltanufnv}
    \end{minipage}%
\end{figure}


\section{CONCLUSION\label{sec:conclusion}}
Using nearly simultaneous radio to NIR to UV data with temporal intervals up to 10 days, we conducted SED studies of the blazar OJ 287. We constructed 106 SEDs covering $MJD$ from 54850 (2009-01-19) to 59227 (2021-01-13) and modeled them in the $\log \nu$-$\log \nu f_{\nu}$ diagram using a LP synchrotron model. The main results are summarized as follows:
\begin{enumerate}
    \item All the constructed 106 SEDs can be well fit by the LP model, with curvature $b > 0.02$. The $b$ values ranged from $0.038$ to $0.234$, with a median of $0.117\pm0.045$.
    \item We classified the observational periods into flare and quiescent segments based on whether the flux values at $V$ band exceed or fall below $10^{-10.49}\ \rm erg\ cm^{-2}\ s^{-1}$, the mean plus half the standard deviation of the $V$-band flux distribution. We found that the median flux at peak frequency of the SEDs during flare segments was 0.37$\pm$0.22 dex higher than during quiescent segments, while no significant differences were observed in the median values of the curvature parameter $b$ or the peak frequency $\lognup$.  
    \item There is a significant relation between the $V$ band magnitude and $B-V$ color index for the 106 SEDs, confirming a bluer-when-brighter (BWB) relation. A stronger BWB trend is found in the flare segments compared to the quiescent ones, as further supported by the anti-correlation between the SED peak frequency and the $V$ band magnitude. 
    \item We found a significant anti-correlation between the SED curvature $b$ and the peak frequency $\lognup$ of the synchrotron component. The slope of the correlation between $1/b$ and $\lognup$, measured as 5.79, aligns more closely with the prediction of the statistical acceleration scenario than with the stochastic acceleration scenario, though a notable discrepancy persists. This discrepancy indicates that additional factors, such as deviations from idealized conditions or radiative contributions—such as thermal emission from the accretion disk in the optical-UV range during quiescent states—may play a role in producing the observed steeper slope.
    \item Within the framework of the statistical acceleration mechanism, we considered potential factors influencing the observed SED changes in blazars. No positive correlation was found between changes in peak intensity and peak frequency, suggesting that change in electron energy distribution is unlikely to be the primary driver. Other factors, such as changes in Doppler boosting factor and/or magnetic fields, may contribute to the observed SED changes.
\end{enumerate}

\section*{ACKNOWLEDGMENTS}
We thank the referee for insightful suggestions, which have significantly improved the draft.
We thank M.S. Anjum for helpful discussion of the physical mechanism for LP SEDs. 
WWZ is supported by the science research grants from the China Manned Space Project with No. CMSCSST-2021-A06. 
ACG is partially supported by Chinese Academy of Sciences (CAS) President's International Fellowship Initiative (PIFI) (grant no. 2016VMB073). 
MFG is supported by the National Science Foundation of China (grant 12473019), Shanghai Pilot Program for Basic Research-Chinese Academy of Science, Shanghai Branch (JCYJ-SHFY-2021-013), the National SKA Program of China (Grant No. 2022SKA0120102), the science research grants from the China Manned Space Project with No. CMSCSST-2021-A06. 
S.K. was funded by the European Union ERC-2022-STG - BOOTES - 101076343. Views and opinions expressed are however those of the author(s) only and do not necessarily reflect those of the European Union or the European Research Council Executive Agency. Neither the European Union nor the granting authority can be held responsible for them. 
P.K. acknowledges support from the Department of Science and Technology (DST), Government of India, through the DST-INSPIRE faculty grant (DST/ INSPIRE/04/2020/002586).
LC is supported by the National Science Foundation of China (grant 12173066), the National SKA Program of China (Grant No. 2022SKA0120102) and Shanghai Pilot Program for Basic Research-Chinese Academy of Science, Shanghai Branch (JCYJ-SHFY-2021-013). 
QY is supported by the National Key R\&D Intergovernmental Cooperation Program of China (2022YFE0133700), the Regional Collaborative Innovation Project of Xinjiang Uyghur Autonomous Region (2022E01013), the National Natural Science Foundation of China (12173078).

The research at Boston University was supported in part by National Science Foundation grant AST-2108622, and several NASA Fermi Guest Investigator grants; the latest is 80NSSC23K1508. 
The work at UMRAO was supported in part by a series of grants from the NSF and NASA, most recently AST-0607523 and NASA Fermi GI grants NNX09AU16G, NNX10AP16G, NNX11AO13G, and NNX13AP18G. 
This research has made use of data from the OVRO 40-m monitoring program supported by private funding from the California Insitute of Technology and the Max Planck Institute for Radio Astronomy, and by NASA grants NNX08AW31G, NNX11A043G, and NNX14AQ89G and NSF grants AST-0808050 and AST-1109911. 
This publication makes use of data obtained at Mets\"ahovi Radio Observatory, operated by Aalto University in Finland. The various diligent observers of Aalto University in Finland are thankfully acknowledged. The Very Long Baseline Array (VLBA) is an instrument of the National Radio Astronomy Observatory. The National Radio Astronomy Observatory is a facility of the National Science Foundation operated by Associated Universities, Inc.. This study is based in part on observations conducted using the Perkins Telescope Observatory (PTO) in Arizona, USA, which is owned and operated by Boston University. This paper has also made use of up-to-date SMARTS optical/near-infrared light curves that are available at \url{www.astro.yale.edu/smarts/glast/home.php}. SMARTS observations of Large Area Telescope-monitored blazars are supported by Yale University and Fermi GI grant NNX 12AP15G, and the SMARTS 1.3-m observing queue received support from NSF grant AST-0707627. Data from the Steward Observatory spectropolarimetric monitoring project were used. This program is supported by Fermi Guest Investigator grants NNX08AW56G, NNX09AU10G, NNX12AO93G and NNX15AU81G.  

\bibliography{ms}
\bibliographystyle{apj}

\end{document}